\documentclass[12pt]{iopart}
\usepackage{iopams}
\usepackage{epsf}
\usepackage{latexsym}
\usepackage{epsfig}
\usepackage{amssymb}
\usepackage{amsfonts}
\usepackage{amssymb}
\usepackage{graphicx}

\newcommand{\mb}[1]{{\mathbf{#1}}}
\newcommand{\gb}[1]{\mbox{\boldmath{$#1$}}}
\newcommand{\prs}[1]{{\left(#1\right)}}
\newcommand{\chs}[1]{{\left\{#1\right\}}}
\newcommand{\col}[1]{{\left[#1\right]}}
\newcommand{\norm}{\mathcal{N}}
\newcommand{\prob}{\mathcal{P}}
\newcommand{\parf}{\mathcal{Z}_n}

\newcommand{\icn}[2]{{<#1_1\cdots #1_#2>}}

\newcommand{\sgn}{{\mbox{sgn}\,}}
\newcommand{\pder}[2]{\frac{\partial #1}{\partial #2}}
\newcommand{\avg}[2]{{\left<#1\right>_{#2}}}
\newcommand{\iav}{{ A_1,...,A_L,\mb{r},\mb{t}}}
\newcommand{\tavg}[1]{\avg{#1}{\iav}}

\newcommand{\fdv}[2]{{\frac{\delta #1}{\delta #2}}}
\newcommand{\hpi}{{\hat{\pi}}}
\newcommand{\hx}{{\hat{x}}}
\newcommand{\hy}{{\hat{y}}}
\newcommand{\hq}{{\hat{q}}}
\newcommand{\hQ}{{\hat{Q}}}

\newcommand{\tf}{{\tilde{f}}}
\newcommand{\bx}{{\mathbf{x}}}
\newcommand{\bhx}{{\mathbf{\hx}}}
\newcommand{\atanh}{\mbox{atanh\,}}

\newcommand{\cut}[1]{{}}

\begin{document}

\title[Statistical Mechanics of MIMO Channels]{Statistical Mechanics
Analysis of LDPC Coding in MIMO Gaussian Channels}
\author{Roberto C.~Alamino and David Saad}
\address{Neural Computing Research Group, Aston University, Birmingham B4 7ET, UK}

\begin{abstract}
Using analytical methods of statistical mechanics, we analyse the
typical behaviour of a multiple-input multiple-output (MIMO)
Gaussian channel with binary inputs under LDPC network coding and
\emph{joint decoding}. The saddle point equations for the replica
symmetric solution are found in particular realizations of this
channel, including a small and large number of transmitters and
receivers. In particular, we examine the cases of a single
transmitter, a single receiver and the symmetric and asymmetric
interference channels. Both dynamical and thermodynamical
transitions from the ferromagnetic solution of perfect decoding to
a non-ferromagnetic solution are identified for the cases
considered, marking the practical and theoretical limits of the
system under the current coding scheme. Numerical results are
provided, showing the typical level of improvement/deterioration
achieved with respect to the single transmitter/receiver result,
for the various cases.
\end{abstract}

\maketitle

\section{Introduction}

The statistical physics of disordered systems has been
systematically developed over the past few decades to analyse
systems of interacting components under different interaction
regimes~\cite{Nishibook,MPV}. It enables one to derive typical
macroscopic properties of systems comprising a large number of
units under conditions of quenched disorder, which correspond to
different randomly sampled instances of the problem.

While their origin lies in the study of spin
glasses~\cite{Edwards75,SK,SKPRB}, methods of statistical
mechanics have been successfully employed to study a broad range
of interdisciplinary subjects, from thermodynamics of fluids to
biological and even sociological problems. In these studies, the
problems were mapped onto known statistical physics models, such
as Ising spin systems, and analysed using established methods
and techniques from statistical physics.

In particular, these methods have been successfully employed
recently to investigate hard computational
problems~\cite{monasson2,Hartmann05} as well as problems in information
theory~\cite{ksRev} and multi-user
communication~\cite{tanakaCDMA}. They proved to be highly useful
for gaining insight into the properties of the problems studied
and in providing exact typical case results that complement the
rigorous bounds reported in the theoretical computer science and
information theory literature.

In the current study we employ the powerful analytical methods of
statistical mechanics to examine the typical properties of
Multiple-Input Multiple-Output (MIMO) communication channels where
messages are encoded using state of the art Low-Density
Parity-Check (LDPC) error correcting
codes~\cite{gallager,richardson,urbanke,mackaybook}.

MIMO channels are becoming increasingly more relevant in modern
communication networks that rely on adaptive and ad-hoc
configurations. Sensor networks, for instance, may rely on
simultaneous transmission of information from a large number of
transmitters that give rise to high levels of interference; while
multiple access, at various levels, is exercised daily by millions
of mobile phone users.

This problem of communication over a MIMO channel is particularly
amenable to a statistical physics based analysis for the following
reasons: Firstly, previous studies in the areas of LDPC
error-correcting codes~\cite{ksRev} and Code Division Multiple
Access (CDMA)~\cite{tanakaCDMA,TS03} paved the way for the study
of MIMO systems; and secondly, the framework of multi-user
communication channels is difficult to analyse using traditional
methods of information theory~\cite{cover}, but can be readily
accommodated within the statistical physics framework,
particularly in the case of a large number of users.

The paper is organised as follows: In section~\ref{sec:framework} we
introduce the model to be analysed, followed by statistical physics
framework in section~\ref{sRAD}. We then study several communication
channels: a single transmitter and multiple receivers in
section~\ref{sOSMR}, multiple access in~\ref{sMAC} and symmetric and
asymmetric interference channels in section~\ref{sIC}. In each of the
sections we will consider both cases of a small and large number of
users. We conclude with general insights and future directions.

%
%
\section{The Model}
\label{sec:framework}

As the communication model considered is based on LDPC codes we
will briefly introduce their main characteristics, within the
single channel setting, before describing the MIMO communication
channel to be studied.

LDPC codes, introduced originally by Gallager~\cite{gallager}, are
used to encode $N$-dimensional message vectors $\mb{s}$ into $M$-dimensional codewords $\mb{t}$.
They are defined by a binary
matrix $A=[C_1 \mid C_2]$, called \emph{parity-check matrix}, concatenating two very sparse matrices
known to both sender and receiver, with $C_2$ (of dimensionality
$(M-N)\times(M-N)$) being invertible and $C_1$ of dimensionality
$(M-N)\times N$. The matrix $A$ can be either random or
structured, characterised by the number of non-zero elements per
row/column. Irregular codes show superior performance with respect
to regular constructions~\cite{richardson,idosaad1} if they are
constructed carefully. However, to simplify the presentation, we
focus here on regular constructions; the generalisation of the
methods presented here to irregular constructions is
straightforward~\cite{LDPCchapter,VSK}.

Encoding refers to the mapping of a $N$-dimensional binary vector
$\mb{s} \in \{0,1\}^N$ (original message) to $M$-dimensional
codewords
 $\mb{t}\in\{0,1\}^M$ ($M>N$) by the linear product
\begin{equation}
\label{eq:encoding} \mb{t} = G \mb{s} \ \ \mbox{(mod 2)} \ ,
\end{equation}
where all operations are performed in the field $\{0,1\}$ and are
indicated by $\mbox{(mod 2)}$. The generator matrix is of the form
\begin{equation}
\label{eq:generator} G = \left[ \begin{array}{c} I \\  C_{2}^{-1}
C_{1} \end{array} \right] \ \ \mbox{(mod 2)} \ ,
\end{equation}
where $I$ is the $N\times N$ identity matrix. By construction $A G
= 0 \ \mbox{ (mod 2)}$ and the first $N$ bits of $\mb{t}$
correspond to the original message $\mb{s}$.

Decoding is carried out by estimating the most
probable transmitted vector from the received corrupted
codeword~\cite{LDPCchapter,ksRev}.

In this work, we analyse a MIMO Gaussian channel with $L$ sender
and $O$ receiver units. In this channel, $L$ original binary
messages $s_i\in \{0,1\}^N$, $i=1,...,L$ are encoded using LDPC
error-correcting codes with \emph{independently chosen}
parity-check matrices $A_i$ for each message into binary codewords
$t_i\in \{0,1\}^M$.

Note that, both messages $s_i$ and codewords $t_i$ are vectors and
should include two different indices, the bit index and a separate
index for the number of senders/receivers ($i$). For brevity, we will
reserve the boldface notation for denoting the sets in the
sender/receiver indices and will explicitly denote the bit
index. \cut{For example, we will use $t_i\equiv(t_i^1,...t_i^M)$ and
$\mb{t}^\mu\equiv(t_1^\mu,...,t_L^\mu)$ with $\mu=1,...,M$.}

We concentrate here on regular
Gallager codes, with exactly $K$ non-zero elements per row and $C$
non-zero elements per column in the parity check matrix, which
obey the relation $C=(1-R)K$, where $R=N/M$ is the code rate. The
codewords are transmitted in discrete units of time. \cut{Clearly,
the work can be easily extended to irregular codes that generally
show improved performance over regular
constructions~\cite{richardson,idosaad1}.}

In order to apply the tools of statistical mechanics, we use, for
mathematical convenience, the transformation
\begin{equation}
  x\rightarrow (-1)^x,
\end{equation}
to map the Boolean variables $t_i\in\{0,1\}^M$ onto spin variables
$t_i\in\{1,-1\}^M$. Although they are different variables, we
denote both with the same letter $t_i$. The appropriate use of
each one of them will be clear from the context. At each discrete
time step $\mu$, the (already mapped) vector
$\mb{t}^\mu$, $\mu=1,...,M$ is transmitted
and corrupted by additive white Gaussian noise (AWGN)  obeying the
equation
\begin{equation}
  \mb{r}^\mu =  S \mb{t}^\mu +\gb{\nu}^\mu,
\end{equation}
where $S$ is an $O\times L$ matrix with elements $S_{ji}$, and the
Gaussian noise, independent of the time, is given by the vector
$\gb{\nu}^\mu=(\nu_1^\mu,...,\nu_O^\mu)$ with $\nu_j^\mu
\sim\norm(0,\sigma_j^2)$, $j=1,...,O, \, \forall\mu$, i.e., of zero
mean and variance $\sigma_j^2$.

The matrix $S$, which we call the \emph{interference matrix},
plays an essential role in the current analysis as it crosses
messages between senders and receivers and is responsible for
important interference effects.

%
%
\section{Replica Analysis}
\label{sRAD}

The statistical mechanics based analysis focuses on the decoding
process as it is directly linked to the Hamiltonian within the
physics framework~\cite{sourlas94}.

Decoding is carried out along the same lines as in LDPC error-correcting codes;
the estimate of the first $N$ bits of the
codeword, which contain the original uncoded message, will be made
by introducing $L$ dynamical variable values $\tau_i\in\{\pm1\}^M$,
representing candidate vectors for each of the transmitted codewords.
These will eventually give rise to the estimate of the various
codewords $\left\{ t_i \right\}, ~i=1,\ldots L$, by the $O$
receivers, each of which has access to all the received messages.

In the statistical analysis, we are interested in the behaviour
averaged over the system's disorder, given by the quenched
variables $\mb{r}$, all possible encodings (or equivalently, all
parity-check matrices $A_i$, for each sender) and all transmitted
codewords $t_i$.

If we allow some degree of error in the decoding, in the form of a
prior error probability, the estimator which minimises the bit
error probability is the Marginal Posterior Maximiser (MPM) for
each dynamical variable~\cite{iba,VSK}.
\begin{equation}
  \hat{t_i}^\mu= \sgn\avg{\tau_i^\mu}{\prob(\gb{\tau}|\mb{r})},
\end{equation}
where $\gb{\tau}=(\tau_1,...,\tau_L)$.

The expected overlap between the estimated and the transmitted
codewords serves as a quality measure for the error correction
performance
\begin{equation}
  d_i = \frac1M \sum_{\mu=1}^M \avg{t_i^\mu\sgn\avg{\tau_i^\mu}{\prob(\gb{\tau}|\mb{r})}}{\iav},
\end{equation}
where the average is taken over the joint probability distribution
$\prob(A_1,...,A_L,\mb{r},\mb{t})$. These performance measures
will also be indicative of the dynamical transition from the
ferromagnetic solution of perfect decoding to a non-ferromagnetic
solution as reflected in their values. Note that the receiver
will only get all the messages correctly if $d_i=1$ for all $i$ at
the same time.

The free-energy in the thermodynamic limit $M\rightarrow \infty$ is given by
\begin{equation}
  f=-\lim_{M\rightarrow \infty} \frac1{\beta M L} \avg{\ln Z}{\iav},
\end{equation}
where $Z$ is the partition function
\[
  Z = \sum_{\gb{\tau}} \exp \col{-\beta\sum_{j=1}^O
      \mathcal{H}_j(\gb{\tau}|\mb{r})},
\]
with the Hamiltonian component for each receiver $j$
\begin{equation}
  \mathcal{H}_j(\gb{\tau}|\mb{r})=\frac1{2\sigma_j^2} \sum_{\mu=1}^M
  \prs{r_j^\mu-\sum_{i=1}^L S_{ji}\tau_i^\mu}^2.
\end{equation}
The Hamiltonian gives rise to a likelihood term for the agreement
between the received aggregated vector and the candidate
codewords. The decoding temperature $\beta$ is considered the same
for every receiver and each $\tau_i$ obeys the parity-check
constraint, which for the spin variables is defined by
\begin{equation}
  \prod_{\mu=1}^M (\tau_i^\mu)^{(A_i)_{\nu\mu}}=1, \qquad \nu=1,...,M-N.
\end{equation}

The decoding process is aimed at maximising the probability
\begin{equation}
  \prob(\gb{\tau}|\mb{r})=\frac1Z \exp\col{-\beta\sum_{j=1}^O \mathcal{H}_j(\gb{\tau}|\mb{r})},
\end{equation}

To calculate $\left< \ln Z\right>_\iav$ in the thermodynamic
limit, where $M,N\rightarrow\infty$ while keeping the code rate
$R=N/M$ constant, we use the replica method~\cite{Nishibook,MPV}
which relies on the identity
\begin{equation}
  \left<\ln Z\right> = \lim_{n\rightarrow0}\pder{\ln\left<Z^n\right>}{n},
\end{equation}
and employs an analytical continuation of integer values of $n$ to
a real value that approaches zero. The calculations will follow
the same guidelines as in~\cite{VSK} and we refer the reader to
the appendix for further details.

The partition function is given by
\begin{equation}
  \label{fPF}
  Z  =\sum_{\gb{\tau}} \col{\prod_{i=1}^L\chi (A_i,\tau_i)} \exp\col{-\beta\sum_{j=1}^O\sum_{\mu=1}^M
       \frac1{2\sigma_j^2} \left(r_j^\mu-\sum_{i=1}^L S_{ji} \tau_i^\mu\right)^2},
\end{equation}
where $\chi$ is an indicator function, which is zero if $\tau_i$
does not obey the parity-check equations defined by the matrix $A_i$.

We assume that the matrices $A_i$ are chosen from the same
ensemble of parity-check matrices, which means that all code rates
will be the same, $R_i=R$. From information theoretical
considerations, the capacity region is then given by
\begin{equation}
  \alpha R<\mathcal{C},
\end{equation}
where $\alpha\equiv L/O$ is a characteristic constant of the system called its \emph{load} and
$\mathcal{C}$, the capacity with joint decoding for an
arbitrary distribution of inputs, is obtained by conventional
information theoretical methods~\cite{cover} and is given by
\begin{equation}
  \label{fCCD}
  \mathcal{C}=\frac12 \log_2 \det(I_O+SS^T C_\nu^{-1}),
\end{equation}
where $T$ indicates transposition, $I_O$ is the $O$-dimensional
unit matrix and $C_\nu$ is an $O$-dimensional square diagonal
noise matrix given by $(C_\nu)_{jk} = \sigma_j^2 \delta_{jk}$. This
result will be used as a benchmark and an upper bound for our
results

%
%
\section{Single Transmitter}
\label{sOSMR}

In this and the following sections we compare the replica symmetric results with the known information theoretical
limits. The case $L=O=1$ is easily seen to recover the usual results for a simple Gaussian channel as obtained
in~\cite{VSK}. In the particular case of one sender and an arbitrary number of receivers, the channel matrix is an
$O$-dimensional column vector. The Replica Symmetric (RS) saddle point equations are (see~\ref{sA_OS})
\begin{eqnarray}
  \label{fRSL1O2}
  \hpi(\hx) &= \avg{\delta\prs{\hx-\prod_{l=1}^{K-1}x^l}}{\bx},\\
  \pi(x)    &= \avg{\delta\prs{x-\tanh\col{\sum_{l=1}^{C-1}\atanh \hx^l+\beta\sum_{j=1}^O
               \frac{r_jS_j}{\sigma_j^2}}}}{\bhx,r},
\end{eqnarray}
with
\begin{equation}
  r \sim \prod_{j=1}^O \norm\prs{S_j,\sigma_j^2},
\end{equation}
and where the averages $\avg{}{\bx}$ and $\avg{}{\bhx}$ are taken with
respect to the distributions $\pi(x)$ and $\hpi(\hx)$, respectively.

The overlap is given by
\begin{eqnarray}
  \label{overlap}
  d           &= \avg{\sgn(\rho)}{\rho},   \mbox{~~~with} \\
  \prob(\rho) &= \avg{\delta\prs{\rho-\tanh\col{\sum_{l=1}^C\atanh \hx^l+\beta\sum_{j=1}^O
                  \frac{r_jS_j}{\sigma_j^2}}}}{\bhx,r}.
\end{eqnarray}

The free-energy is
\begin{eqnarray}
  \label{freeeng}
 &\beta f =  \frac{C}K \ln 2 +C\avg{\ln(1+x\hx)}{x,\hx}
             -\frac{C}K\avg{\ln\prs{1+\prod_{m=1}^K x^m}}{\bx}\nonumber\\
           & -\avg{\ln\chs{\sum_{\tau=\pm 1}\exp\col{-\sum_{j=1}^O \frac{\beta}{2\sigma_j^2}
             \prs{r_j-S_j\tau}^2}\prod_{l=1}^C\prs{1+\tau\hx^l}}}{\bhx,r}.
\end{eqnarray}

The ferromagnetic solution,
\begin{eqnarray}
  \label{fFMS}
  \hpi(\hx)=\delta(\hx-1),  \mbox{~~~and~~~}\pi(x)=\delta(x-1),
\end{eqnarray}
represents perfect decoding; it is always present for all noise
levels and has free-energy $f=O/2$.

The internal energy and the entropy can be derived from the free energy by the well-known relations
\begin{equation}
  \label{eng_entropy}
  u=\frac{\partial}{\partial\beta}(\beta f), \qquad s=\beta(u-f).
\end{equation}

Let us study the symmetric case where all transmitters emit with the
same unit power, all entries of $S$ are equal to 1, and all receivers
experience the same noise level $\sigma^2$. The capacity, as given by
equation (\ref{fCCD}), is
\begin{equation}
  \label{fCapL1O2}
  \mathcal{C}=\frac12 \log_2 \prs{1+\frac{O}{\sigma^2}}.
\end{equation}
and the Shannon limit of perfect decoding is attained when $R=\mathcal{C}$, giving for the threshold noise the result
\begin{equation}
  \label{fTL1}
  \sigma^2=\frac{O}{2^{2R/O}-1}.
\end{equation}

To obtain numerical solutions for the various cases we iterated the saddle-point
equations~(\ref{fRSL1O2}), using population
dynamics, and then calculated the quantities of interest such as
the overlap $d$, the free energy $f$ and the entropy $s$ of
equations~(\ref{overlap}(-(\ref{eng_entropy}).

\begin{figure}
\centering
\includegraphics[width=11cm]{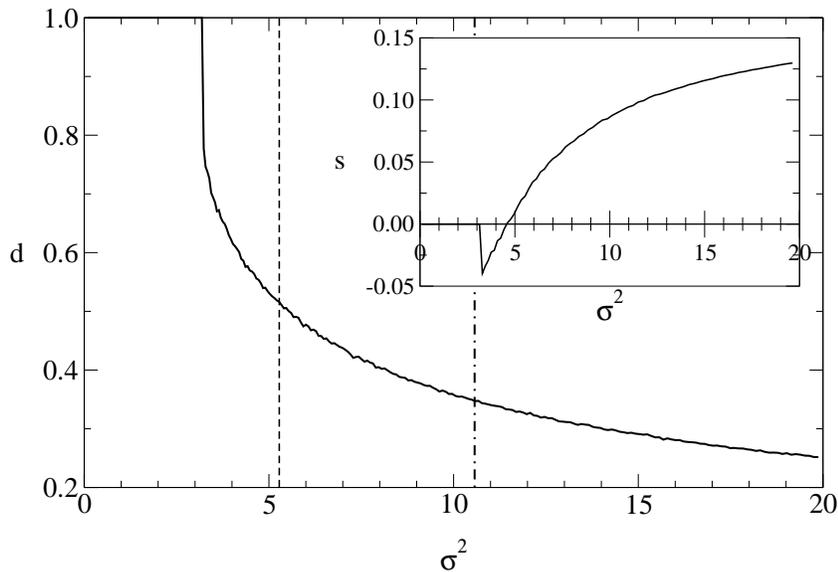}
\caption{Overlap in the single-sender case for $O=2$. The solid line
         describes the result obtained by iterating the saddle point
         equations~(\ref{fRSL1O2}) from arbitrary initial
         conditions. The dotted-dashed line shows the theoretical
         limit obtained from equation~(\ref{fTL1}) and the dashed line
         shows the theoretical limit for sending a doubled message via
         a single Gaussian channel. The inset shows a plot of the
         entropy; the point where the entropy becomes negative marks
         the emergence of metastable states and the dynamical
         transition point, while the point where it crosses back the
         zero entropy line marks the thermodynamical transition noise
         value.}
\label{fgL1O2}
\end{figure}

Figure~\ref{fgL1O2} shows the overlap for $L=1$ (one sender), $O=2$
(two receivers), $\sigma_j^2=\sigma^2$ (equal noise level for all
receivers), $S_j=1$ and $R=1/4$ (with $K=4$ and $C=3$) at the
Nishimori temperature $\beta=1$. The choice of the Nishimori
temperature simplifies the analysis as it is known that for this
temperature, the system does not enter the spin-glass
phase~\cite{Nishibook}. Similar to the case of LDPC codes, there is no
difference between the RS results, obtained using the Nishimori
condition, and those obtained using the replica symmetry breaking
ansatz for the noisy channel studied here~\cite{franzLDPC,Migliorini};
this motivates our present choice of the replica symmetric ansatz.

We can see that the overlap has the value 1 up to the noise level
termed the \emph{dynamical transition} point. This means that while
the noise level is kept below this point, all the receivers can
perfectly recover the transmitted message as the ferromagnetic
solution is the only stable solution. The ferromagnetic solution
remains dominant between this point and the \emph{thermodynamical
transition point}, which marks the noise level where the
non-ferromagnetic state becomes dominant; although an exponential
number of sub-optimal stable solutions in this range prevent the
iterative population dynamics from converging to the ferromagnetic
solution (starting from an arbitrary initial state).

The entropy plot in the inset clarifies the type of solutions
obtained as the noise level increases: the entropy is zero up to
the dynamical transition, meaning that the only stable state is
the ferromagnetic one. Metastable suboptimal solutions emerge
above this point which could be explored using the replica
symmetry breaking ansatz~\cite{franzLDPC,Migliorini}; these
contribute to (unphysical) negative entropy values in this
range~\cite{VSK}. The point where the entropy line crosses the
coordinate axis coincides with the \emph{thermodynamical transition} point.
The thermodynamical transition is always upper bounded by the
Shannon theoretical limit, which is also shown in the overlap plot
as a vertical dashed line.

In table~\ref{tL1} we compare the theoretical limit of sending the
same message $O$ times via a simple Gaussian channel (one sender and
one receiver) with noise level equal to the one considered here
(second column) with the theoretical limit for the MIMO channel given
by equation~(\ref{fTL1}) (third column) and the points of the
dynamical (fourth column) and the thermodynamical (fifth column)
transitions obtained by numerical integration of the RS equations for
$O=1,2,3$ receivers. It is clear that the
dynamical and thermodynamical transitions occur always before the
theoretical limit. As expected, the more receivers are added, the
higher the noise level the system can tolerate. However, the
differences between the dynamical and the thermodynamical transition
values, and between the thermodynamical transition and theoretical
limit increase. Both are related to the fact that, in adding more
receivers, we also increase the number of metastable states in the
system; these emerge earlier and contribute to a higher entropy.

Comparing the theoretical noise limit for sending the message $O$
times by a simple Gaussian channel with the limit for the MIMO channel
with one sender and $O$ receivers, we can see that the later is just
$O$ times the former. This can be understood noting that the
information being sent in the MIMO channel is the same as in the
$O$-replicated Gaussian channel, but with $O$ times the power; while
in the MIMO channel case, the $O$ bits are sent with power 1 at each
time step. We can see by that the results of the RS ansatz that the
transition points are even below the theoretical limit for the simple
Gaussian channel and significantly below the MIMO limit. This clearly
shows that in this type of communication channel, even with joint
decoding, the available information is being poorly used. It makes a
strong case for the use of \emph{network coding}, i.e., to encode
jointly the vectors $\mb{t}^\mu$ prior to transmission.

Network coding, for instance using fountain codes~\cite{LT,tornado},
is likely to make a better use of the resource by generating codewords
that are more suited for better extraction of information under joint
decoding.

For $O>3$ the numerical instabilities grow larger with $O$ and a precise evaluation of the
points is increasingly more difficult.

\begin{table}[ht]
  \centering
  \caption{Comparison between the Shannon limit for a simple Gaussian
           channel and the MIMO channel, the dynamical transition
           point and the thermodynamical transition for the
           single-sender case ($L=1$).}
  \begin{tabular}{ccccc}
    \hline\hline
    $O$ & Shannon's Limit   & Shannon's Limit &Dynamical& Thermodynamical \\
      & (Gaussian Channel)& (MIMO Channel) & Transition& Transition\\
    \hline
    1 & 2.41 & 2.41 & 1.59 & 2.24 \\
    2 & 5.28 & 10.57 & 3.28 & 4.59 \\
    3 & 8.17 & 24.50 & 4.90 & 6.68 \\
    \hline
  \end{tabular}
  \label{tL1}
\end{table}

Another case of interest is that of an infinite number of receivers.
When $O\rightarrow\infty$, the average over the $r$ variables in
equation~(\ref{fRSL1O2}), can be substituted by an average over the
Gaussian variable
\begin{equation}
  v \equiv\sum_{j=1}^O\frac{r_jS_j}{\sigma_j^2}~.
\end{equation}
In the case, of equal noise and $S_j=1$, this variable has zero mean
and variance $O/\sigma^2$ which reflects the signal to noise ratio
appearing in the capacity expression~(\ref{fCapL1O2}).

%
%
\section{Multiple Access Channel}
\label{sMAC}

The multiple access channel (MAC) is a particular case where
$O=1$ and $S$ is an $L$-dimensional row matrix. Let us consider
once more the symmetric case where $S_{ji}=1$ and
$\sigma^2_j=\sigma^2$. The capacity then becomes
\begin{equation}
  \mathcal{C}=\frac12 \log_2 \prs{1+\frac{L}{\sigma^2}},
\end{equation}
and the threshold noise
\begin{equation}
  \label{fTL2}
  \sigma^2=\frac{L}{2^{2LR}-1}.
\end{equation}

In this case, due to the interference effect in the received
message, one should guarantee that the interference term has the
correct order with respect to $L$. Taking into account that the
received messages are independent random variables, we normalise
their sum by the factor of $1/\sqrt{L}$.

The simplest case is $L=2$ and the RS saddle point equations for user 1 are given by
\begin{eqnarray}
  \label{fRSL2O1}
  &\hpi_1(\hx_1) = \avg{\delta\prs{\hx-\prod_{l=1}^{K-1}x_1^l}}{\bx},\\
  &\pi_1(x_1)    = \left<\delta\left(x-\tanh\left\{\sum_{l=1}^{C-1}\atanh \hx_1^l+\frac{\beta r}{\sigma^2\sqrt{2}}
                   \right.\right.\right.\nonumber\\
                &  \left.\left.\left.+\frac12\ln\col{\frac{1-\tanh\prs{\frac{\beta}{2\sigma^2}}
                   \tanh\prs{\sum_{l=1}^C\atanh \hx_2^l+\frac{\beta r}{\sigma^2\sqrt{2}}}}
                   {1+\tanh\prs{\frac{\beta}{2\sigma^2}}\tanh\prs{\sum_{l=1}^C\atanh
           \hx_2^l+\frac{\beta r}{\sigma^2\sqrt{2}}}}}\right\}\right)\right>_{\bhx,r},
\end{eqnarray}
and the overlap is
\begin{eqnarray}
  &d_1           = \avg{\sgn(\rho)}{\rho},\\
  &\prob(\rho)   = \left<\delta\left(\rho-\tanh\left\{\sum_{l=1}^C\atanh \hx_1^l+\frac{\beta r}{\sigma^2\sqrt{2}}
                   \right.\right.\right.\nonumber\\
                &  \left.\left.\left.+\frac12\ln\col{\frac{1-\tanh\prs{\frac{\beta}{2\sigma^2}}
                   \tanh\prs{\sum_{l=1}^C\atanh \hx_2^l+\frac{\beta r}{\sigma^2\sqrt{2}}}}
                   {1+\tanh\prs{\frac{\beta}{2\sigma^2}}\tanh\prs{\sum_{l=1}^C\atanh \hx_2^l+\frac{\beta r}{\sigma^2\sqrt{2}}}}}
                   \right\}\right)\right>_{\bhx,r},
\label{overlap1}
\end{eqnarray}
where
\begin{equation}
  r \sim \norm\prs{\sqrt{2},\sigma^2}.
\end{equation}

The corresponding equations for user 2 are identical to
(\ref{fRSL2O1})-(\ref{overlap1}) except for interchanging the
indices 1 and 2. The free-energy is given by
\begin{eqnarray}
 &\beta f =  \frac{C}K \ln 2 +\frac{C}2\sum_{i=1}^2\avg{\ln(1+x_i\hx_i)}{x,\hx}
             -\frac{C}{2K}\sum_{i=1}^2\avg{\ln\prs{1+\prod_{m=1}^K x_i^m}}{\bx}\nonumber\\
           & - \frac12\avg{\ln\chs{\sum_{\tau_1,\tau_2}\exp\col{-\frac{\beta}{2\sigma^2}
               \prs{r-\frac{\tau_1+\tau_2}{\sqrt{2}}}^2}
               \prod_{i=1}^2\prod_{l=1}^C\prs{1+\tau_i\hx_i^l}}}{\bhx,r}~.
\end{eqnarray}
For the ferromagnetic solution equation~(\ref{fFMS}) results in $f = 0.25$.
Indeed, for the MIMO Gaussian channel studied in this paper, we always have that the ferromagnetic free energy is given by
$f = 1/2\alpha$.

\begin{figure}
\centering
\includegraphics[width=11cm]{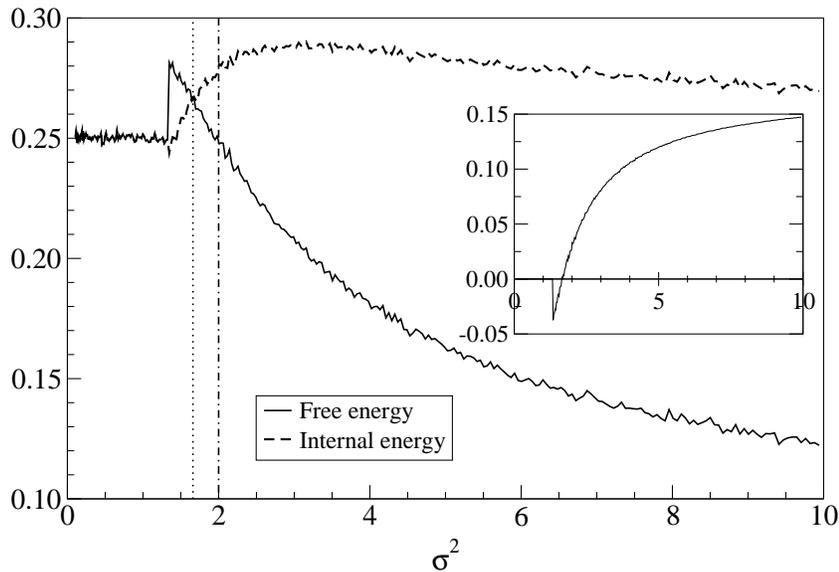}
\caption{Free energy and internal energy in the MAC case for $L=2$
represented by the solid and dashed lines, respectively; the results
have been obtained by iterating the saddle point
equations~(\ref{fRSL1O2}) from arbitrary initial conditions. The
dot-dashed line shows the theoretical limit obtained from
equation~(\ref{fTL2}) and the dotted line the thermodynamical
transition point.  The entropy as a function of the noise level is
shown in the inset; the point where the entropy becomes negative marks
the emergence of metastable states and the dynamical transition point,
while the point where it crosses back the zero entropy line marks the
thermodynamical transition noise value.}
\label{fgL2O1}
\end{figure}

By iteratively solving the saddle point equations we obtain quantities
of interest for this case. The free and internal energies, for $L=2$
and $R=1/4$ ($K=4$ and $C=3$) and at the Nishimori temperature, are
represented in figure~\ref{fgL2O1} by the solid and dashed lines,
respectively; Shannon's theoretical threshold is given by
$\sigma^2=2$, indicated by the dot-dashed line.  The point where the
free energy differs from the internal energy, which
corresponds to the overlap changing from 1 to lower values, marks the
dynamical transition point. The thermodynamical transition point is
identified by the crossing of the two energies and is denoted by the
dotted line. The entropy function, shown in the inset plotted against
the noise level, also helps to identify the dynamical and
thermodynamical transition points (where the entropy becomes negative
and where it crosses back the coordinate axis, respectively). Both
points are below Shannon's limit.

Table~\ref{tMAC} shows the results for $L=1,2,3$ senders. The second
column gives the theoretical limit obtained from equation
(\ref{fCCD}); it shows the deterioration in performance as the number
of senders increases. The deterioration is also evident in the results
obtained by numerical results obtained using the RS ansatz given by
the dynamical transition (second column) and the thermodynamical
transition (third column). Contrary to the single-sender case the
difference between the transition points decreases with
increasing~$L$; this reflect the fact that additional inputs seem to
increase the number sub-optimal solution states (and hence reduce
their free energy and affect the thermodynamical transition point) but
have a lesser effect on the onset of the metastable states.

\begin{table}[ht]
  \centering
  \caption{Comparison between the Shannon's limit, the dynamical transition point and the thermodynamical
           transition for the MAC case ($O=1$).}
  \begin{tabular}{cccc}
    \hline\hline
    L & Shannon's & Dynamical  & Thermodynamical \\
      & Limit     & Transition & Transition \\
    \hline
    1 & 2.41 & 1.59 & 2.24 \\
    2 & 2.00 & 1.32 & 1.66 \\
    3 & 1.64 & 1.24 & 1.45 \\
    \hline
  \end{tabular}
  \label{tMAC}
\end{table}

There are two possible scenarios one may consider in the case of a
large number of users ($L\rightarrow\infty$). The first is the random
interference scenario. Due to the well-known isomorphism between CDMA
and MIMO channels, this case is exactly the one calculated
in~\cite{tanakaCDMA} if one rescales $S_{ji}=s_{ji}/\sqrt{L}$ where
the $s_{ji}$ are i.i.d. random variables with zero mean, unit variance
and vanishing odd moments. The second scenario is the deterministic
interference case, where the matrix $S$ is not random.  This scenario
is of little interest as the capacity grows with the logarithm of the
number of users while the transmitted information grows linearly with
the number of transmitters; the capacity per user goes to zero in this
limit, rendering the communication infeasible.

%
%
\section{Interference Channel}
\label{sIC}

\emph{Interference channel}~\cite{cover} refers to a scenario where
several transmitters send data simultaneously to an equal number of
receivers; the transmission from a given transmitter to the
corresponding receiver is corrupted by (small) interference from all
other transmitters. The receivers can then communicate with each other
to optimally extract the original messages. Some sensor networks are
among the most well known exemplars of systems that could be modelled
by an interference channel.

\begin{figure}
\centering
\includegraphics[width=11cm]{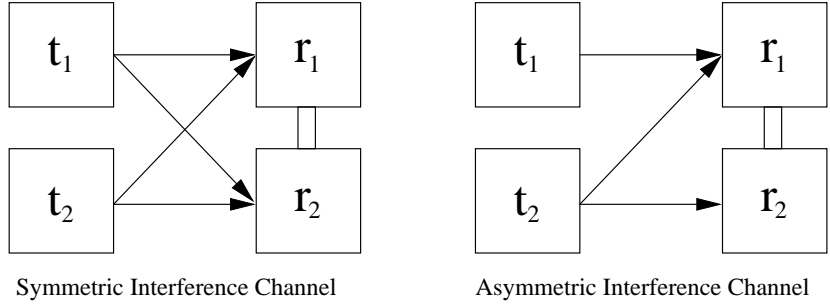}
\caption{Diagram representing the symmetric (left) and asymmetric
         (right) interference channels.  The first and second
         transmitters and receivers are denoted by $t_1$, $t_2$ and
         $r_1$, $r_2$, respectively. Arrows represent the transmitted
         messages and the double line bewtween the receivers indicates
         \emph{joint decoding}.}
\label{fig:diagram}
\end{figure}

In the following, we will study two basic types of interference
channels, the symmetric and the asymmetric case. For simplicity, we
limit the number of transmitters and receivers considered here to
$L=O=2$; this will make the interpretation of the results easier and
more transparent.  Both channels are depicted in
figure~\ref{fig:diagram}.  The symmetric case corresponds to the
transmitters sending messages to both receivers (left picture) while
in the asymmetric case only the first transmitter sends a message to
the first receiver while the second transmitter sends a message to
both receivers.

\subsection{The Symmetric Interference Channel}
\label{subsection:SIC}

We first study the case $L=O=2$ with a symmetric interference matrix
\begin{equation}
  S=\left(
       \begin{array}{cc}
     1 & \epsilon \\
     \epsilon & 1
       \end{array}
    \right),
\end{equation}
where $0<\epsilon\leq1$. The corresponding capacity can be derived
using equation~(\ref{fCCD}) to obtain
\begin{equation}
\label{capacity_interference}
  \mathcal{C}=\frac12 \log_2 \col{1+\frac{2(1+\epsilon^2)}{\sigma^2}+\frac{(1-\epsilon^2)^2}{\sigma^4}}.
\end{equation}

The RS saddle point equations are given by
\begin{eqnarray}
\label{saddlepointInterference}
  &\hpi_1(\hx_1)  =\avg{\delta\prs{\hx_1-\prod_{l=1}^{K-1}x_1^l}}{\bx},\\
  &\pi_1 (x_1)    =\left< \delta\left(x_1-\tanh\left\{\sum_{l=1}^{C-1}\atanh \hx_1^l+
                   \frac{\beta}{\sigma^2\sqrt{2}}(r_1+\epsilon r_2)\right.\right.\right.\nonumber\\
                &  +\left.\left.\left.
                   \frac12 \ln \col{\frac
                   {1-\tanh(\frac{\beta\epsilon}{\sigma^2})\tanh\prs{\frac{\beta(\epsilon r_1+r_2)}{\sigma^2\sqrt{2}}
                   + \sum_{l=1}^C\atanh\hx_2^l}}
                   {1+\tanh(\frac{\beta\epsilon}{\sigma^2})\tanh\prs{\frac{\beta(\epsilon r_1+r_2)}{\sigma^2\sqrt{2}}
                   + \sum_{l=1}^C\atanh\hx_2^l}}
                   }\right\}\right)\right>_{\bhx,r},
\end{eqnarray}
where
\begin{equation}
  r_i\sim\norm\prs{\frac{1+\epsilon}{\sqrt{2}},\sigma^2}, \qquad i=1,2.
\end{equation}
The corresponding equations for $\hpi_2$ and $\pi_2$ are similar to those of
$\hpi_1$ and $\pi_1$ and can be obtained by interchanging the indices 1 and 2.

Note that the same scaling as in the MAC case is necessary here due to
the interference. However, for $\epsilon=0$, this scaling should be
omitted as the interference vanishes, leaving two separate Gaussian
channels.

The overlaps are given by
\begin{eqnarray}
  &d_i  =\avg{\sgn(\rho)}{\rho}, \\
  &\prob(\rho)   = \left< \delta\left(\rho-\tanh\left\{\sum_{l=1}^C\atanh \hx_1^l+
                   \frac{\beta}{\sigma^2\sqrt{2}}(r_1+\epsilon r_2)\right.\right.\right.\nonumber\\
                 &  +\left.\left.\left.
                    \frac12 \ln \col{\frac
                   {1-\tanh(\frac{\beta\epsilon}{\sigma^2})\tanh\prs{\frac{\beta(\epsilon r_1+r_2)}{\sigma^2\sqrt{2}}
                   + \sum_{l=1}^C\atanh\hx_2^l}}
                   {1+\tanh(\frac{\beta\epsilon}{\sigma^2})\tanh\prs{\frac{\beta(\epsilon r_1+r_2)}{\sigma^2\sqrt{2}}
                   + \sum_{l=1}^C\atanh\hx_2^l}}
                   }\right\}\right)\right>_{\bhx,r}.
\end{eqnarray}

The free-energy $f$ is
\begin{eqnarray}
 &\beta f =  \frac{C}K \ln 2 +\frac{C}2\sum_{i=1}^2\avg{\ln(1+x_i\hx_i)}{x,\hx}
             -\frac{C}{2K}\sum_{i=1}^2\avg{\ln\prs{1+\prod_{m=1}^K x_i^m}}{\bx}\nonumber\\
           & - \frac12\left<\ln\left\{\sum_{\tau_1,\tau_2}\exp\col{
             -\frac{\beta}{2\sigma^2}\prs{r_1-\frac{\tau_1+\epsilon\tau_2}{\sqrt{2}}}^2
             -\frac{\beta}{2\sigma^2}\prs{r_2-\frac{\epsilon\tau_1+\tau_2}{\sqrt{2}}}^2}\right.\right.\nonumber\\
       & \left.\left.\times\prod_{i=1}^2\prod_{l=1}^C\prs{1+\tau_i\hx_i^l}\right\}\right>_{\bhx,r}.
\end{eqnarray}

Accordingly, the free-energy of the ferromagnetic solution
(\ref{fFMS}) is $f=0.5$, as for the simple Gaussian channel.

We solved numerically the saddle point
equations~(\ref{saddlepointInterference}) and calculated
quantities of relevance in this case. The graphs for the overlap,
entropy and energy are qualitatively the same as in the other two
cases, with a similar behaviour with the appearance of both
dynamical and thermodynamical transition points before Shannon's
limit.

\begin{figure}
\centering
\includegraphics[width=11cm]{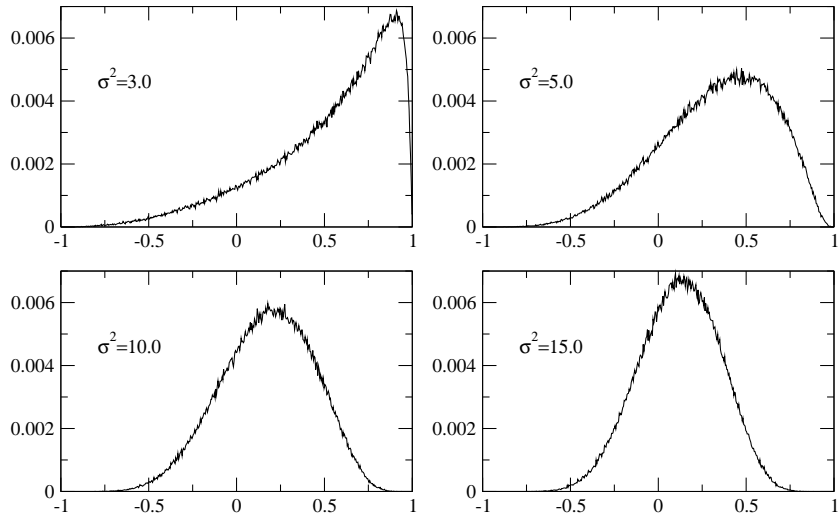}
\caption{Profile of the $\pi$ distribution. The plots are
histograms with 500 bins for a population of 40000 fields.
         The noise level value $\sigma^2$ is indicated in each graph.
         All noise levels are above
     the dynamical transition point; below the transition point the distribution is a
     delta function $\delta(x-1)$.
     Note how the distribution changes slowly from $\delta(x-1)$
     to $\delta(x)$ as the noise level increases.}
\label{fgL2O2a}
\end{figure}

\begin{figure}
\centering
\includegraphics[width=11cm]{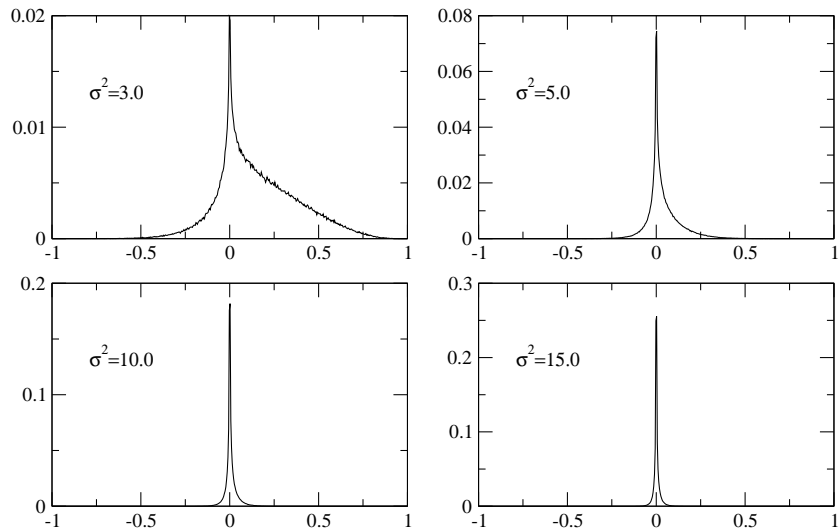}
\caption{Profile of the $\hpi$ distribution. The plots are
histograms with 500 bins for a population of 40000 fields.
         The noise level value $\sigma^2$ is indicated in each graph. All noise levels are above
     the dynamical transition point; below it the profile distribution is simply a
     delta function $\delta(\hx-1)$. In this case, the profile changes abruptly from $\delta(\hx-1)$
       to an asymmetric distribution centred at $\hx=0$ and diverges rapidly to $\delta(\hx)$ with
      the  increasing noise.}
\label{fgL2O2b}
\end{figure}

Figures~\ref{fgL2O2a} and \ref{fgL2O2b} show the field
distributions $\pi(x)$ and $\hpi(\hx)$, respectively, for four
different values of the noise level in the RS ansatz; with
$\epsilon=1.0$, $\beta=1$ and $R=1/4$ ($K=4$, $C=3$). It should be
noticed that, before the dynamical transition point, these
distributions are delta functions centred at 1, corresponding to
the ferromagnetic solution~(\ref{fFMS}). The plotted distributions
are histograms with 500 bins for 40000 fields. In
figure~\ref{fgL2O2a} we see how the $\pi$ distribution changes
slowly from the delta function in 1 to a delta function in zero,
which is the solution for $\sigma^2\rightarrow\infty$. The $\hpi$
distribution depicted in figure~\ref{fgL2O2b} changes abruptly
from the delta function in 1 to
a highly peaked asymmetric distribution around zero (paramagnetic
solution) when the dynamical transition point is crossed. Looking at
the values on each of the graphs, it is visible how the scales
increase very fast as the noise level attains higher values.

\begin{figure}
\centering
\includegraphics[width=11cm]{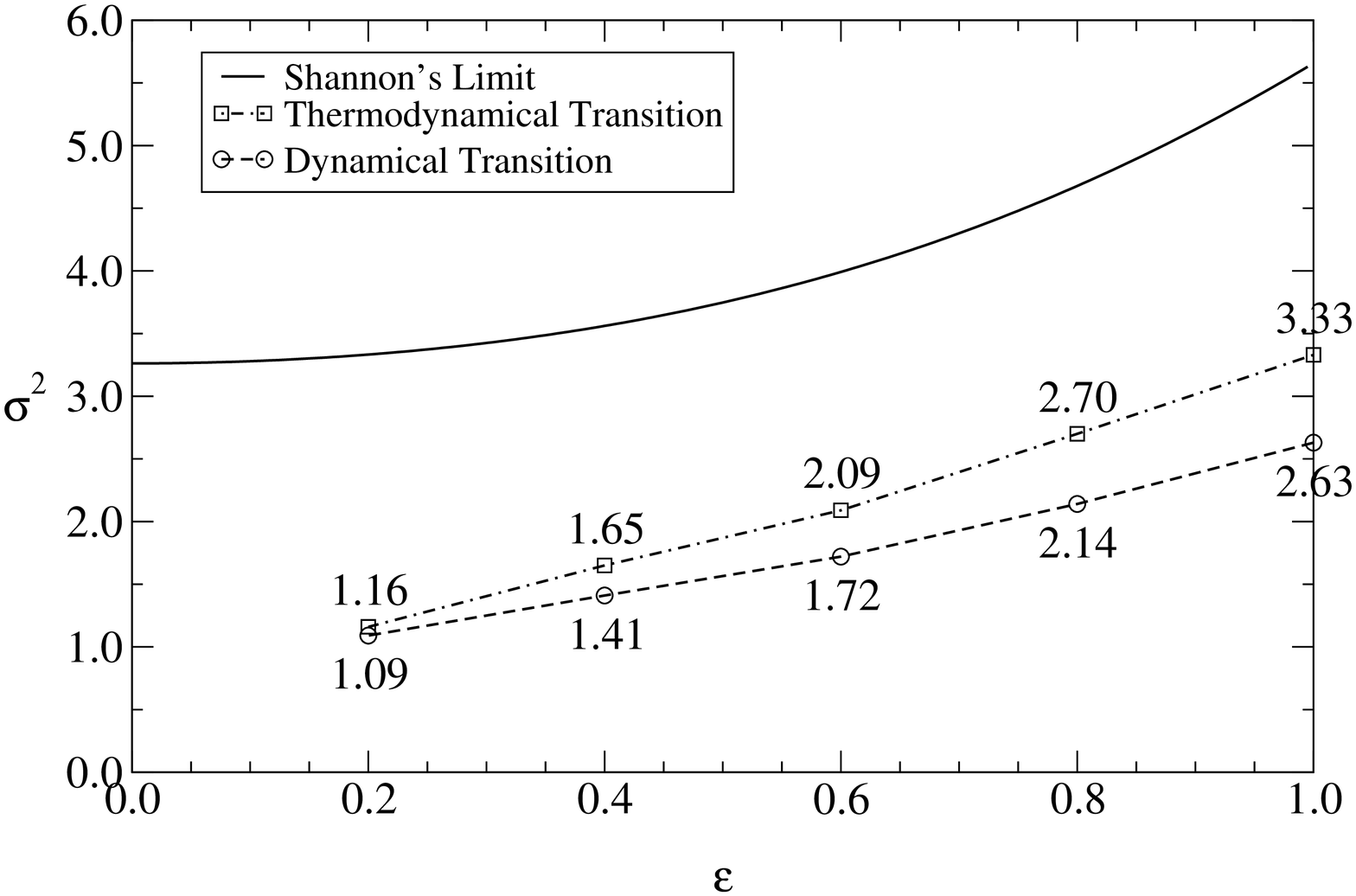}
\caption{Transition points and theoretical limits as a function of
the interference level $\epsilon$. The solid line represents the
theoretical limit obtained from information theoretical methods;
the dashed-dotted and dashed lines correspond to the thermodynamical
and dynamical transition points, respectively.} \label{fgL2O2Cap}
\end{figure}

If one keep a constant code rate $R=1/4$ but allows $\epsilon$ to vary, one obtains the dependence of the
threshold noise as a function of $\epsilon$, depicted in
figure~\ref{fgL2O2Cap} (for $\beta=1$, $K=4$ and $C=3$). Both
dynamical (dashed line) and thermodynamical transition values
(dashed-dotted line) are upper bounded by the theoretical
limit. Although this may seem counterintuitive, the communication
resilience against noise increases with the interference level. This
can be understood in the case of joint detection by noting that the
increased interference provides more information about the other
transmitters, such that higher levels of noise can be tolerated by
joint decoding.

For large $O$ with $L\sim\mathcal{O}(1)$ or large $L$ with
$O\sim\mathcal{O}(1)$, the results should approach those obtained for
large number of users in the single transmitter and in the MAC case,
respectively. The behaviour must be dictated by the value of the
system load $\alpha$. In this case, we expect the results to cross
from a behaviour similar to the one of a MAC channel for $\alpha>1$ to
one that resembles the single transmitter case for $\alpha<1$. We are
currently working on the analytical and computational aspects of this
last case as well as on the case of large $O$ and $L$ values while
keeping the ration $L/O \sim \mathcal{O}(1)$ finite.

\subsection{The Asymmetric Interference Channel}
\label{subsection:AIC}

A variant of the interference channel discussed in
section~\ref{subsection:SIC}, for the case of $L=O=2$, is the
asymmetric interference channel. This realisation of the
interference channel is highly relevant to cases where receivers
are distributed at random and experience different noise levels,
for instance, in the case of sensor networks. The interference
matrix is asymmetric in this case and takes the form (for $L=O=2$)
\begin{equation}
  S=\left(
       \begin{array}{cc}
     1 & \epsilon \\
     0& 1
       \end{array}
    \right),
\end{equation}
with $0<\epsilon\leq1$. The corresponding capacity is now (again by (\ref{fCCD}))
\begin{equation}
\label{cap_asym_int}
  \mathcal{C}=\frac12 \log_2 \col{1+\frac{(2+\epsilon^2)}{\sigma^2}+\frac1{\sigma^4}}.
\end{equation}

The RS saddle point equations are given by
\begin{eqnarray}
\label{sadptAsymInt}
  &\hpi_i(\hx_i)  =\avg{\delta\prs{\hx_i-\prod_{l=1}^{K-1}x_i^l}}{\bx}, \qquad i=1,2,\\
  &\pi_1 (x_1)    =\left< \delta\left(x_1-\tanh\left\{\sum_{l=1}^{C-1}\atanh \hx_1^l+
                   \frac{\beta r_1}{\sigma^2\sqrt{2}}\right.\right.\right.\nonumber\\
                &  +\left.\left.\left.
                   \frac12 \ln \col{\frac
                   {1-\tanh(\frac{\beta\epsilon}{2\sigma^2})\tanh\prs{\frac{\beta\epsilon r_1}{\sigma^2\sqrt{2}}
                   +\frac{\beta r_2}{\sigma^2}+ \sum_{l=1}^C\atanh\hx_2^l}}
                   {1+\tanh(\frac{\beta\epsilon}{2\sigma^2})\tanh\prs{\frac{\beta\epsilon r_1}{\sigma^2\sqrt{2}}
                   +\frac{\beta r_2}{\sigma^2}+ \sum_{l=1}^C\atanh\hx_2^l}}
                   }\right\}\right)\right>_{\bhx,r},\\
  &\pi_2 (x_2)    =\left< \delta\left(x_2-\tanh\left\{\sum_{l=1}^{C-1}\atanh \hx_2^l+
                   \frac{\beta\epsilon r_1}{\sigma^2\sqrt{2}}+\frac{\beta r_2}{\sigma^2}
                   \right.\right.\right.\nonumber\\
                &  +\left.\left.\left.
                   \frac12 \ln \col{\frac
                   {1-\tanh(\frac{\beta\epsilon}{2\sigma^2})\tanh\prs{\frac{\beta r_1}{\sigma^2\sqrt{2}}
                   + \sum_{l=1}^C\atanh\hx_1^l}}
                   {1+\tanh(\frac{\beta\epsilon}{2\sigma^2})\tanh\prs{\frac{\beta r_1}{\sigma^2\sqrt{2}}
                   + \sum_{l=1}^C\atanh\hx_1^l}}
                   }\right\}\right)\right>_{\bhx,r},
\end{eqnarray}
where
\begin{equation}
  r_1\sim\norm\prs{\frac{1+\epsilon}{\sqrt{2}},\sigma^2}, \qquad r_2\sim\norm\prs{1,\sigma^2}.
\end{equation}

In this case, the scaling $1/\sqrt{L}$ (in this case $L=2$,
although the treatment can be extended to include a general number
of sources) appears only in the first receiver, as it is being
affected by the interference.

The overlaps are given by
\begin{eqnarray}
  &d_i           = \avg{\sgn(\rho_i)}{\rho_i}, \qquad i=1,2,\\
  &\prob(\rho_1) = \left<\delta\left(\rho_1- \tanh\left\{\sum_{l=1}^C\atanh \hx_1^l+
                   \frac{\beta r_1}{\sigma^2\sqrt{2}}\right.\right.\right.\nonumber\\
  &                +\left.\left.\left.
                   \frac12 \ln \col{\frac
                   {1-\tanh(\frac{\beta\epsilon}{2\sigma^2})\tanh\prs{\frac{\beta\epsilon r_1}{\sigma^2\sqrt{2}}
                   +\frac{\beta r_2}{\sigma^2}+ \sum_{l=1}^C\atanh\hx_2^l}}
                   {1+\tanh(\frac{\beta\epsilon}{2\sigma^2})\tanh\prs{\frac{\beta\epsilon r_1}{\sigma^2\sqrt{2}}
                   +\frac{\beta r_2}{\sigma^2}+ \sum_{l=1}^C\atanh\hx_2^l}}
                   }\right\}\right)\right>_{\bhx,r},\\
  &\prob(\rho_2) = \left< \delta\left(\rho_2-\tanh\left\{\sum_{l=1}^C\atanh \hx_2^l+
                   \frac{\beta\epsilon r_1}{\sigma^2\sqrt{2}}+\frac{\beta r_2}{\sigma^2}
                   \right.\right.\right.\nonumber\\
                &  +\left.\left.\left.
                   \frac12 \ln \col{\frac
                   {1-\tanh(\frac{\beta\epsilon}{2\sigma^2})\tanh\prs{\frac{\beta r_1}{\sigma^2\sqrt{2}}
                   + \sum_{l=1}^C\atanh\hx_1^l}}
                   {1+\tanh(\frac{\beta\epsilon}{2\sigma^2})\tanh\prs{\frac{\beta r_1}{\sigma^2\sqrt{2}}
                   + \sum_{l=1}^C\atanh\hx_1^l}}
                   }\right\}\right)\right>_{\bhx,r},
\end{eqnarray}

The free-energy $f$ is obtained from
\begin{eqnarray}
 &\beta f =  \frac{C}K \ln 2 +\frac{C}2\sum_{i=1}^2\avg{\ln(1+x_i\hx_i)}{x,\hx}
             -\frac{C}{2K}\sum_{i=1}^2\avg{\ln\prs{1+\prod_{m=1}^K x_i^m}}{\bx}\nonumber\\
           & - \frac12\left<\ln\left\{\sum_{\tau_1,\tau_2}\exp\col{
             -\frac{\beta}{2\sigma^2}\prs{r_1-\frac{\tau_1+\epsilon\tau_2}{\sqrt{2}}}^2
             -\frac{\beta}{2\sigma^2}\prs{r_2-\tau_2}^2}\right.\right.\nonumber\\
       & \left.\left.\times\prod_{i=1}^2\prod_{l=1}^C\prs{1+\tau_i\hx_i^l}\right\}\right>_{\bhx,r},
\end{eqnarray}
with the free energy of the ferromagnetic solution $f=0.5$.

\begin{figure}
\centering
\includegraphics[width=14cm]{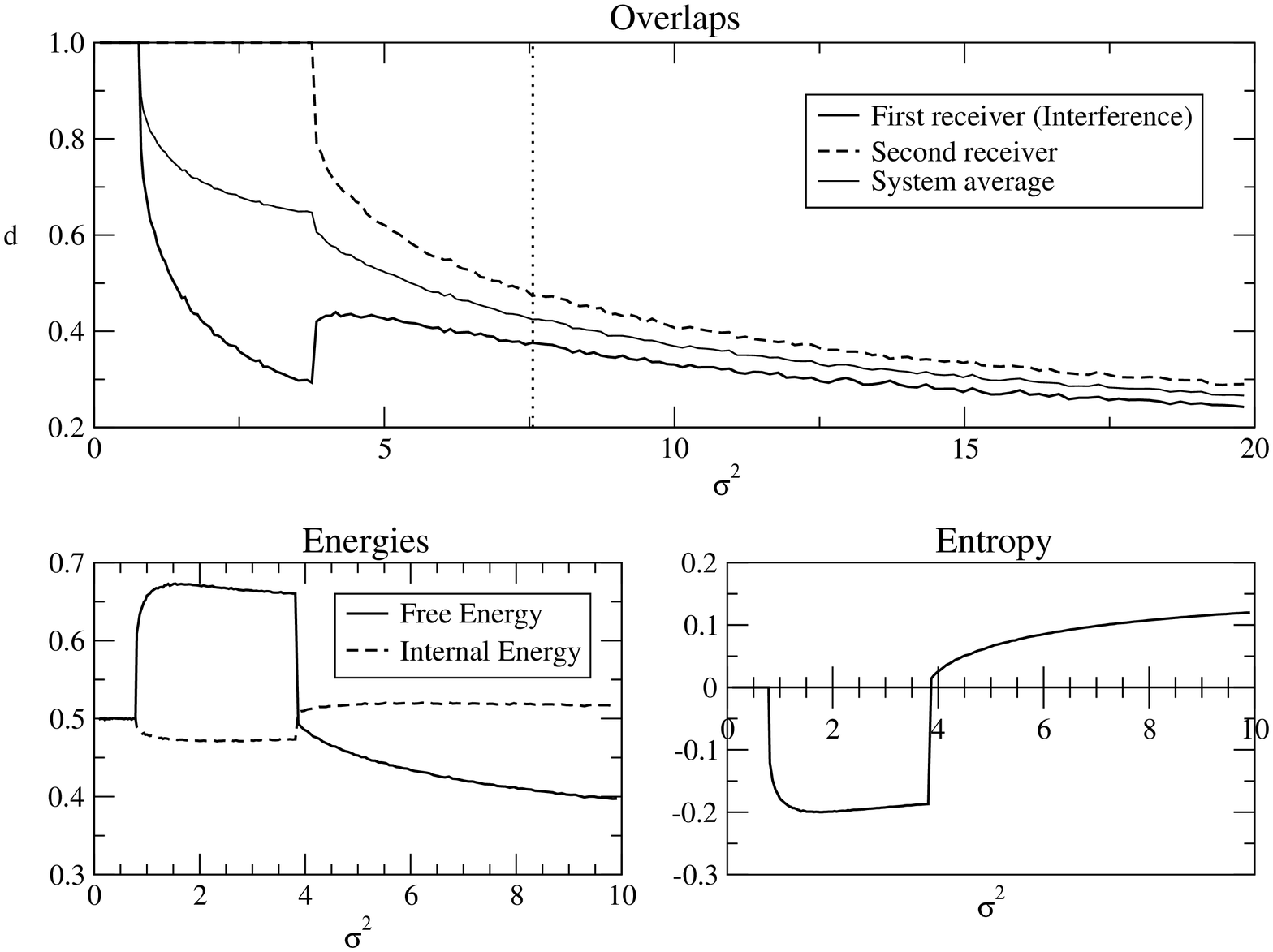}
\caption{Numerical integration of saddle point equations for the
         asymmetric interference channel. The upper plot shows the
         overlap for the first receiver (thick continuous line), which
         suffer the effects of interference, the second receiver
         (thick dashed line) and the average overlap for the entire
         system (thin continuous line). The Shannon limit for the
         system is depicted by the dotted vertical line. At the
     bottom, the left graph shows the free-energy (continuous
         line) and the internal energy (dashed line) for the entire
         systems while the right graph shows the entropy values
         obtained under the RS ansatz.}
\label{fgL2O2c}
\end{figure}

The numerical solution of~(\ref{sadptAsymInt}) with $\epsilon=1.0$,
$\beta=1$ and $R=1/4$ ($K=4$, $C=3$) leads to the results depicted in
figure \ref{fgL2O2c}. The top plot shows the overlaps of the solutions
obtained for both receivers.  Interestingly, the overlaps for the two
receivers behave significantly differently in spite of the fact that
messages are decoded jointly; this one of the striking features of the
asymmetric interference channel. The overlap for both receivers is one
up to the point where the first receiver (thick continuous line),
which experiences interference effects, exhibits a dynamical
transition which signals the practical noise threshold for this
system. The same point can be identified in the entropy plot (bottom
right) as the point where the entropy becomes negative. This point is
very far from Shannon's limit (dotted line) $\sigma^2\approx7.56$;
this can be explained by the additional metastable states introduced
by the asymmetric interference. Note that the first receiver undergoes
a (dynamical) transition before the second receiver (thick dashed
line) that does not suffer from interference; this is in spite of the
fact that the messages are decoded jointly.

However, the dynamical transition point for the second receiver
introduces an unexpected behaviour at the first receiver. When the
overlap for the second receiver drops to sub-optimal levels, the first
receiver exhibits a sudden {\em increase} in its decoding overlap.
This behaviour may be understood by examining the average overlap,
which can be viewed as the overlap for the entire system. We see that
the system's overlap suffers a second transition at this point,
although the average overlap continues to decrease monotonically; the
{\em system as a whole} has a certain amount of retrievable
information which keeps decreasing with the noise level.

The above result shows that the way information is distributed among
the receivers can be highly non-trivial. It also shows that for
systems with many users, the thermodynamical transition point is
determined mostly by the weakest node (which experiences the highest
levels of interference) and may lead practical limits that are very
far from the Shannon bound.

%
%
\section{Conclusions}
\label{sC}

We investigated the properties of coded Gaussian MIMO
channels using methods of statistical mechanics. The problems
investigated relate to the cases of a single transmitter, multiple
access and interference in the case of multiple receivers and
transmitters. In all cases, transmissions are coded using LDPC
error-correcting codes.

The method used in the analysis, the replica approach, enables one
to obtain typical case results that complements the theoretical
bounds reported in the information theory literature. The
numerical results obtained for particular MIMO channels and
parameter values are presented and contrasted with the
information theoretical results.

MIMO channels are characterised by an interference matrix $S$
which mixes inputs from the various transmitters to provide the
messages at the receiving end. We examine cases where the
interference matrix is deterministic.
This requires the
introduction of a non-trivial scaling in order to obtain
meaningful results.

The results obtained provide characteristic, typical case, results in
all cases. For the single transmitter and MAC problems we show both
dynamical and thermodynamical transition points as functions
of the number of receivers and transmitters, respectively. We see that
the gaps between the practical and theoretical thresholds (dynamical
and thermodynamical transitions, respectively), and the gap between
them and Shannon's limit, increase with the number of receivers in the
single transmitter case and decrease with the number of transmitters
in the MAC case.

In the single transmitter case, this results from the increase in the
number of variables and consequentially the increase in the number of
metastable states. The point where metastable solutions emerge
determines the dynamical transition point (practical threshold), while
the number of metastable states
affects the thermodynamic transition point.  The increasing number of
transmitters in the MAC case enables one to effectively reduce the
noise level by averaging over a higher number of random and
independent noise sources.

The comparison with theoretical limits for the single transmitter
case reveals an important feature of multiuser channels as to how
the available information is used. The huge gap between the
transition points and Shannon's limit is indicative of a poor use
of resource, and suggests network coding as a measure to achieve a
good use of resource; without it, the system's efficiency remains
below the achievable theoretical limit for sending the same
message repeatedly via a simple Gaussian channel. One possible
solution that we are currently investigating is the use of
fountain codes~\cite{LT,tornado} for making a more efficient use
of the available resource.

The main result in the symmetric interference channel case is the increase
in both dynamical and thermodynamical transition points as a
function of the interference parameter $\epsilon$. Results for low
$\epsilon$ values are  similar to the case of separate Gaussian
channels; as $\epsilon$ increases, both values come closer to
Shannon's theoretical limit with the thermodynamical transition
point showing a stronger increase. This could be explained by the
increase of (mixed) information in comparison to the noise level;
this information can be decoded jointly, with an effectively lower
noise level. The more moderate increase in the practical threshold
(dynamical transition) is due to the difficulty in jointly
decoding the various sources in practice due to the emergence of
metastable states.

In the asymmetric case we found a striking different behaviour of
the system. The new feature observed is the second transition
suffered by the system as a whole. We also detected a surprising
behaviour of the receiver which experiences interference; in spite
of the joint decoding, the information available to it is
suppressed by the second receiver. Only when the second receiver
stops decoding perfectly, the performance of the first receiver
improves.

An interesting extension, of significant practical relevance, would be
the extend the LDPC coding framework to complex MIMO channels, where
circular noise is considered~\cite{Moustakas03}. Another as well as
the possible extension is the case of a large number of senders and
receivers where the ration between them remains finite. The study of
these and other related problems is underway.

%
%
\section*{Acknowledgements}

Support from EVERGROW, IP No.~1935 in FP6 of the EU is gratefully
acknowledged.

%
%
\section*{References}

\bibliographystyle{nature}
\bibliography{statmechecc}

%
%
\appendix

\section{Replica Symmetric Calculations}

>From the partition function (\ref{fPF}), we can write the averaged replicated partition function
$\parf\equiv\tavg{Z^n}$ as
\begin{eqnarray}
  &\parf =  \frac{\lambda^M}{2^{NL}}
            \sum_{\chs{\tau_a}}\int d\mb{r}\,
            \exp\col{-\sum_{j=1}^O\sum_{\mu=1}^M \frac1{2\sigma_j^2}
            \left(r_j^\mu-\sum_{i=1}^L S_{ji}\tau_{i0}^\mu\right)^2}\nonumber\\
         &  \times \exp\col{-\sum_{a=1}^n\sum_{j=1}^O\sum_{\mu=1}^M \frac{\beta}{2\sigma_j^2}
            \left(r_j^\mu-\sum_{i=1}^L S_{ji}\tau_{ia}^\mu\right)^2}\col{\prod_{i=1}^L\Lambda_i(\{\tau_{ia}\})},\\
  &\lambda \equiv \prod_{j=1}^O (2\pi\sigma_j^2)^{-1/2},
\end{eqnarray}
where the multiplicative constants come from the normalisation of the probability distributions in the outside average
and we defined $\tau_{i0}\equiv t_i$.
Following \cite{TS03}, we have
\begin{eqnarray}
  \Lambda_i(\{\tau_{ia}\}) &\equiv  \avg{\prod_{a=0}^n \chi(A_i,\tau_{ia})}{A_i}\nonumber\\
                           &= \frac1{N_A}\oint D\mb{Z}_i\col{\sum_{\omega_i}
                       \prs{\frac1M \sum_\mu Z_i^\mu\tau_{ia^i_1}^\mu\cdots\tau^\mu_{ia^i_{m_i}}}^K}^{M-N},
\end{eqnarray}
where
\begin{equation}
  D\mb{Z}_i\equiv \prs{\frac1{2^{M-N}}}^{n+1} \, \prod_{\mu=1}^M \frac{dZ_i^\mu}{2\pi i}\frac1{(Z_i^\mu)^{C+1}},\qquad
  \omega_i\equiv \icn{a^i}{{m_i}},
\end{equation}
and the variables $m_i$ assume all integer values for the index $i$ from 0 to $n+1$.

Defining
\begin{equation}
  q_{\omega_i}\equiv \frac1M \sum_\mu Z_i^\mu\tau_{ia^i_1}^\mu\cdots\tau^\mu_{ia^i_{m_i}},
\end{equation}
and using integral representations for the delta functions, we can write
\begin{eqnarray}
  \parf & =  2^{-NL}\int \prs{\prod_{i=1}^L\prod_{\omega_i} \frac{dq_{\omega_i}d\hq_{\omega_i}}{2\pi i/M}}
             \col{\prod_{i=1}^L\sum_{\omega_i} \prs{q_{\omega_i}}^K}^{M-N} \nonumber\\
    &    \times \prod_{i=1}^L\exp\prs{-M\sum_{\omega_i} q_{\omega_i}\hq_{\omega_i}}\nonumber\\
    &    \times \sum_{\chs{\tau_a}}\prod_{i=1}^L\col{\oint D \mb{Z}_i\exp\prs{\sum_{\omega_i} \hq_{\omega_i}\sum_\mu
             Z_i^\mu\tau_{ia^i_1}^\mu\cdots\tau^\mu_{ia^i_{m_i}}}}\nonumber \\
    &    \times\lambda^M\int d\mb{r}\,\exp\col{-\sum_{j=1}^O\sum_{\mu=1}^M \frac1{2\sigma_j^2}
        \left(r_j^\mu-\sum_{i=1}^L S_{ji}\tau_{i0}^\mu\right)^2}\nonumber\\
        &    \times\exp\col{-\sum_{a=1}^n\sum_{j=1}^O\sum_{\mu=1}^M \frac{\beta}{2\sigma_j^2}
        \left(r_j^\mu-\sum_{i=1}^L S_{ji}\tau_{ia}^\mu\right)^2}.
\end{eqnarray}

Defining
\begin{equation}
  \prod_{i=1}^L\prod_{\omega_i} \frac{dq_{\omega_i}d\hq_{\omega_i}}{2\pi i/M}\equiv DqD\hq,
\end{equation}
and
\begin{equation}
  \gamma=\frac{2^{-(M-N)(n+1)}2^{-N}}{N_A},
\end{equation}
and integrating over the variables $Z_i^\mu$, the $\mu$ indices
factorise and we obtain
\begin{equation}
  \parf=\int Dq D\hq \, \exp\col{ML\tf(q,\hq)},
\end{equation}
with
\begin{eqnarray}
  \tf(q,\hq)& \equiv\frac1M\ln\gamma+\frac{(1-R)}{L} \sum_{i=1}^L\ln
              \col{\sum_{\omega_i}\prs{q_{\omega_i}}^K}\nonumber\\
            &\  -\frac1L\sum_{i=1}^L \sum_{\omega_i}q_{\omega_i}\hq_{\omega_i}+\frac1L\ln \Phi,
\end{eqnarray}
and
\begin{eqnarray}
  \Phi & \equiv \lambda\int d^L r\, \sum_{\chs{\tau_a}}\col{\prod_{i=1}^L
                \frac1{C!}\prs{\sum_{\omega_i} \hq_{\omega_i}\tau_{ia^i_1}\cdots\tau_{ia^i_{m_i}}}^C}
                \nonumber\\
       & \times \exp\col{-\sum_{j=1}^O \frac1{2\sigma_j^2}\prs{r_j-\sum_{i=1}^L S_{ji}\tau_{i0}}^2}\nonumber\\
       & \times \exp\col{-\sum_{a=1}^n\sum_{j=1}^O \frac{\beta}{2\sigma_j^2}\prs{r_j-\sum_{i=1}^L S_{ji}\tau_{ia}}^2}.
\end{eqnarray}
Using the replica symmetric (RS) ansatz
\begin{eqnarray}
  \label{fRSA}
  q_{\omega_i}   & = q^i_0 \avg{(x_i)^{m_i-\Delta_i}}{x_i},\qquad x_i\sim\pi_i(x_i),\nonumber\\
  \hq_{\omega_i} & = \hq^i_0 \avg{(\hx_i)^{m_i-\Delta_i}}{\hx_i},\qquad\hx_i\sim\hpi_i(\hx_i),
\end{eqnarray}
where
\begin{eqnarray}
  \Delta_i&=\left\{
    \begin{array}{cl}
      1, & 0\in\{a_1^i,...,a_{m_i}^i\} \\
      0, &\mbox{otherwise}.
    \end{array}
  \right.
\end{eqnarray}

For small $n$
\begin{eqnarray}
  \ln\col{\sum_{\omega_i} \prs{q_{\omega_i}}^K} &= \ln\col{2 (q^i_0)^K}+n\avg{\ln\prs{1+\prod_{m=1}^K x_i^m}}{\bx},
\end{eqnarray}
where $\avg{\cdot}{\bx}$ indicates the average over all variables $x_i^m$ and
\begin{equation}
  \sum_{\omega_i} q_{\omega_i} \hq_{\omega_i} = 2 q^i_0\hq^i_0\col{1+n\avg{\ln (1+x_i\hx_i)}{x_i,\hx_i}},
\end{equation}
\begin{equation}
   \sum_{\omega_i} \hq_{\omega_i} \tau_{ia^i_1}^\mu\cdots\tau^\mu_{ia^i_{m_i}} =
   \hq^i_0\prs{1+\tau_{i0}} \avg{\prod_{a=1}^n\prs{1+\tau_{ia}\hx_i}}{\bhx}.
\end{equation}

Inserting the result in $\Phi$ and summing over the zero-th replicas we have
\begin{eqnarray}
  \Phi   = & \frac{\prs{2^L\hQ_0}^C}{(C!)^L}\left< \sum_{\chs{\tau_a}}
         \prod_{l=1}^C\prod_{a=1}^n\prod_{i=1}^L\prs{1+\tau_{ia}\hx_i^l}\right.\nonumber\\
           & \times\left.\exp\col{-\sum_{a=1}^n\sum_{j=1}^O \frac{\beta}{2\sigma_j^2}\prs{r_j-
             \sum_{i=1}^L S_{ji}\tau_{ia}}^2}\right>_{r,\bhx}.
\end{eqnarray}
where $\hQ_0\equiv\prod_i\hq_0^i$ and
\begin{equation}
  r \sim \prod_{j=1}^O \norm\prs{\sum_{i=1}^L S_{ji},\sigma_j^2}.
\end{equation}
The sum over the $n$ replicas factorises to
\begin{eqnarray}
  \label{fPhi}
  &\Phi   = \frac{\prs{2^L\hQ_0}^C}{(C!)^L}\left<\left\{\sum_{\tau_1,...,\tau_L}
            \prod_{l=1}^C\prod_{i=1}^L\prs{1+\tau_i\hx_i^l}\right.\right.\nonumber\\
          & \times\left.\left.\exp\col{-\sum_{j=1}^O \frac{\beta}{2\sigma_j^2}\prs{r_j-\sum_{i=1}^L
            S_{ji}\tau_i}^2}\right\}^n\right>_{r,\bhx}.
\end{eqnarray}

\subsection{Single Transmitter}
\label{sA_OS}

Let us consider $L=1$. Then
\begin{eqnarray}
  & \Phi = \frac{\prs{2\hq_0}^C}{C!}\left<\left\{\sum_{\tau}
             \prod_{l=1}^C\prs{1+\tau\hx^l}\right.\right.\nonumber\\
         & \times\left.\left.\exp\col{-\sum_{j=1}^O \frac{\beta}{2\sigma_j^2}
             \prs{r_j-S_j\tau}^2}\right\}^n\right>_{r,\bhx}.
\end{eqnarray}
and, for small $n$
\begin{eqnarray}
  & \ln \Phi = \ln \frac{(2\hq_0)^C}{C!}\nonumber\\
  & +n\avg{\ln\chs{\sum_\tau
    \prod_{l=1}^C\prs{1+\tau\hx^l}\exp\col{-\sum_{j=1}^O \frac{\beta}{2\sigma_j^2}
    \prs{r_j-S_j\tau}^2}}}{r,\bhx}.
\end{eqnarray}

Derivations with respect to $q_0$ and $\hq_0$ give $2q_0\hq_0=C$
and functional derivatives with respect to $\pi(x)$ and
$\hpi(\hx)$ give equations~(\ref{fRSL1O2}) of section~\ref{sOSMR}.

\subsection{MAC}

In this case, $O=1$,
\begin{eqnarray}
  &\ln \Phi  = \ln\frac{\prs{2^L\hQ_0}^C}{(C!)^L}\nonumber\\
  & + n\avg{\ln\chs{\sum_{\{\tau_i\}}
               \prod_{l=1}^C\prod_{i=1}^L\prs{1+\tau_i\hx_i^l}
        \exp\col{-\frac{\beta}{2\sigma^2}\prs{r-\sum_{i=1}^L S_i\tau_i}^2}}}{r,\bhx},
\end{eqnarray}
and the corresponding extremisation, including the necessary
normalisation, gives equations~(\ref{fRSL2O1}) of
section~\ref{sMAC}.

\subsection{Interference Channel}

The MIMO case with $L=O=2$ can be viewed as an interference
Gaussian channel where the receivers cooperate with each other to
decode the received message. In this case
\begin{eqnarray}
  &\Phi   = \frac{\prs{4\hQ_0}^C}{C!}\left<\left\{\sum_{\tau_1,\tau_2}
            \prod_{l=1}^C\col{\prs{1+\tau_1\hx_1^l}\prs{1+\tau_2\hx_2^l}}\right.\right.\nonumber\\
          & \times\left.\left. e^{-\frac{\beta}{2\sigma^2}\left(r_1-S_{11}\tau_1-S_{12}\tau_2\right)^2}
            e^{-\frac{\beta}{2\sigma^2}\left(r_2-S_{21}\tau_1-S_{22}\tau_2\right)^2}\right\}^n\right>_{r,\bhx}.
\end{eqnarray}

Extremisation with respect to $\pi_i$, $i=1,2$ results in
\begin{equation}
  \hpi_i(\hx_i)=\avg{\delta\prs{\hx_i-\prod_{l=1}^{K-1}x_i^l}}{\mb{x}},
\end{equation}
and the functional derivative with respect to $\hpi_1$ gives
\begin{eqnarray}
  &\fdv{\tf}{\hpi_1(y_1)}  = -nC\avg{\ln\prs{1+\hy_1 x_1}}{\mb{x}}\nonumber\\
                         & +nC\avg{\ln \col{\sum_{\tau_1,\tau_2} P^{\tau_1\tau_2}
                           \prs{1+\tau_1\hy_1}\prod_{l=1}^{C-1}\prs{1+\tau_1\hx_1^l}
                           \prod_{l=1}^C\prs{1+\tau_2\hx_2^l}}}{r,\mb{\hx}},
\end{eqnarray}
where we defined
\begin{equation}
  P^{\tau_1\tau_2} \equiv e^{-\frac{\beta}{2\sigma^2}\left(r_1-S_{11}\tau_1-S_{12}\tau_2\right)^2}
                          e^{-\frac{\beta}{2\sigma^2}\left(r_2-S_{21}\tau_1-S_{22}\tau_2\right)^2}.
\end{equation}

Equating to zero we obtain
\begin{equation}
  \pi_1(x_1)=\avg{\delta\prs{x_1-h_1\prs{r,\mb{\hx}}}}{r,\mb{\hx}},
\end{equation}
where
\begin{eqnarray}
  h_1\prs{r,\mb{\hx}} & \equiv \frac{\sum_{\tau_1,\tau_2} \tau_1 \, P^{\tau_1\tau_2}
                           \prod_{l=1}^{C-1}\prs{1+\tau_1\hx_1^l}
                           \prod_{l=1}^C\prs{1+\tau_2\hx_2^l}}{\sum_{\tau_1,\tau_2} P^{\tau_1\tau_2}
                           \prod_{l=1}^{C-1}\prs{1+\tau_1\hx_1^l}
                           \prod_{l=1}^C\prs{1+\tau_2\hx_2^l}}.
\end{eqnarray}

The final equations with the interference normalisation are
already given in section ~\ref{sIC} for both cases of a symmetric (subsection \ref{subsection:SIC}) and an asymmetric
(subsection \ref{subsection:AIC})
interference channels. These equations can be easily generalised for
any number of $L$ and $O$ values. In numerical calculations, however,
the numerical errors occurring due to the introduction of additional
fields in this direct form are difficult to control. Clever algebraic
manipulations are necessary to keep these errors under control in
order to obtain accurate results.

\end{document}